\documentclass[letterpaper,twocolumn,10pt]{article}

\usepackage[utf8]{inputenc}
\usepackage[T1]{fontenc}
\usepackage{lmodern}
\usepackage[margin=0.75in]{geometry}
\usepackage{booktabs}
\usepackage{array}
\usepackage{amsmath}
\usepackage{amssymb}
\usepackage{xcolor}
\usepackage{listings}
\usepackage{url}
\usepackage[hidelinks]{hyperref}
\usepackage{balance}
\usepackage{caption}
\usepackage{pgfplots}
\pgfplotsset{compat=1.17}
\newcommand{\rone}{R\textsubscript{1}}   % regex scorer (method under test)
\newcommand{\rtwo}{R\textsubscript{2}}   % blind semantic grader
\newcommand{\rthree}{R\textsubscript{3}} % human canonical
\newcolumntype{C}{>{\centering\arraybackslash}p{0.9cm}}

\title{\textbf{Can AI Write Compliant Code, and to What Extent? Evaluating SOC 2 Compliance of Claude Fable 5, Claude Opus 4.8, and Claude Opus 5 Across Four Use Cases}}

\author{
  Iccha Sethi\\
  \textit{Vanta}\\
  \texttt{iccha.sethi@vanta.com}
  \and
  Herman Errico\\
  \textit{Vanta}\\
  \texttt{herman.v@vanta.com}
}

\date{}

\begin{document}
\maketitle
%------------------------------------------------------------------------------
\begin{abstract}
Software teams now delegate production code to language models, including code
that provisions storage, handles credentials, and stores regulated data. We asked two questions about that code: when nobody mentions security,
does a model apply the controls a SOC~2 program expects (encryption, restricted
access, logging, retention), and how much does one sentence naming the standard
change the answer?

We ran a controlled test: three frontier models (Claude Fable~5, Claude
Opus~4.8, Claude Opus~5) across four use cases (an S3 command-line tool, an
authentication service, an RDS Terraform module, and a file-upload handler
holding personal data), each generated twice, once from a neutral task
statement and once from the same statement plus a single SOC~2 sentence. All 24
generations were scored against binary rubrics whose every item maps to a
specific Trust Services Criterion, and every failure and every flagged insecure
act was hand-verified against the code.

Unprompted conformance ranged from 47\% to 88\%, and the level tracked whether a
control is part of how the code is normally written. Password hashing and
\texttt{storage\_encrypted} appear without being asked for; four S3 hardening
calls, a retention policy, and an MFA hook do not. The neutral prompt also
produced working code that carried real vulnerabilities: across the sixteen
original generations, a reachable Werkzeug debugger that permits remote code
execution, an unauthenticated download endpoint, and an endpoint returning every
stored name and email, all three scored clean by our first commission checklist.
A fourth defect in the newest model, an RDS instance that discards its data on
teardown outside production, was scored passing because the setting is computed
by a conditional and never appears as a literal. The single SOC~2 sentence moved
every case to 86--100\%, worth 23 to 50 points, and removed every insecure
construction, though it left a stable set of controls that sit outside the
model's conception of the task, including MFA hooks, cookie flags, and account
lifecycle, which have to be named individually. Model choice mattered least: two
same-generation models finished within one rubric item in all eight comparison
cells, the resolution floor at one generation per cell, and the newer Opus~5
improved on the two weakest cases by less than the sentence was worth. The
pattern-matching scorer a team would use to audit this code is itself
unreliable, disagreeing with semantic grading on 27 of 216 item judgments and
under-crediting well-engineered code in 21 of them, and it passed the RDS defect
above, so it produces false passes as well as false failures and has to be
replaced by semantic checks.
\end{abstract}

%------------------------------------------------------------------------------
\section{Introduction}
\label{sec:intro}

A model that writes an S3 uploader will produce working code, but whether it
produces \emph{conformant} code is a different question, and it is the one a
compliance program has to answer. The controls at issue are unremarkable:
encryption at rest, restricted access, transport security, audit logging,
retention. Their absence produces the most common audit findings, and their
absence in generated code is invisible at authoring time, because a bucket
created without default encryption behaves exactly like one created with it
until somebody looks.

That invisibility is what separates this question from the one prior work has
answered. The literature on AI-generated code security measures whether a model
introduces a vulnerability
\cite{pearce2022asleep,perry2023users,sandoval2023lostatc,khoury2023chatgpt,
fu2025copilot}, and benchmarks have systematized that into
prompt-to-vulnerability datasets
\cite{tony2023llmseceval,siddiq2022securityeval,siddiq2024sallm,
hajipour2024codelmsec,bhatt2023cyberseceval,peng2025cweval}. A control can be
missing without any line of code being wrong. An uploader that never calls
\texttt{put\_bucket\_encryption} contains no defect a test would catch and still
fails an encryption-at-rest control, so a CWE-oriented benchmark will score it
clean.

We set out to measure two things, stated before any code was generated. First,
what a model does with compliance controls when the prompt says nothing about
them, which is how most requests actually arrive. Second, how much of the gap a
single generic instruction recovers, since that is the cheapest intervention
available to a team and the one they will try first.

The design is a factorial: three models, two prompt conditions, four use cases,
24 generations, each scored against a pre-written binary rubric mapped
item-by-item to the Trust Services Criteria (\S\ref{sec:method}). Predictions
for two of the four use cases were registered before those runs existed, and
four further predictions were registered before the third model arm existed. Two
of the seven original predictions were refuted, and both refutations changed
what we believe.

\noindent\textbf{What we found.}
\begin{itemize}
  \item In three of the sixteen original neutral-prompt generations the code was
    actively vulnerable: a reachable debugger permitting remote code execution,
    and two endpoints exposing other users' documents and personal data. Our
    first commission checklist caught none of them
    (\S\ref{sec:res-commissions}).
  \item Unprompted conformance runs 47\% to 88\% and tracks whether a control is
    embedded in the code idiom; the application-versus-infrastructure split did
    not predict it. Declarative Terraform came in 30 points above an imperative
    boto3 script doing comparable work, which refuted our starting hypothesis
    (\S\ref{sec:res-unprompted}).
  \item One sentence naming SOC~2 is worth 23 to 50 points and moves every case
    to 86--100\% (\S\ref{sec:res-directive}), though a stable set survives it,
    including MFA hooks, \texttt{Secure}/\texttt{HttpOnly} cookie flags, and
    account lifecycle, which have to be named individually
    (\S\ref{sec:res-blindspots}).
  \item The pattern-matching scorer a team would use to check this code is
    directionally biased. It under-credited well-engineered code in 21 of 27
    disagreements with semantic grading, and it passed a real defect whose
    non-compliant value is computed by a conditional and never appears as a
    literal, so it produces false passes as well as false failures
    (\S\ref{sec:measurement}).
  \item Model choice produced no detectable difference within a model
    generation, and a small one across generations
    (\S\ref{sec:res-models}).
\end{itemize}

%------------------------------------------------------------------------------
\section{Related Work}
\label{sec:related}

\noindent\textbf{Security of generated code.}
Pearce et al.\ \cite{pearce2022asleep} found roughly 40\% of GitHub Copilot
completions in security-relevant scenarios contained weaknesses, and controlled
user studies established the behavioral counterpart: participants with an AI
assistant wrote less secure code while believing the opposite
\cite{perry2023users}, with a smaller but measurable cost in C
\cite{sandoval2023lostatc}. Audits of ChatGPT output \cite{khoury2023chatgpt}
and of Copilot code in live GitHub projects \cite{fu2025copilot} corroborate the
rates. Benchmarks including LLMSecEval \cite{tony2023llmseceval}, SecurityEval
\cite{siddiq2022securityeval}, SALLM \cite{siddiq2024sallm}, CodeLMSec
\cite{hajipour2024codelmsec}, CyberSecEval \cite{bhatt2023cyberseceval}, and
CWEval \cite{peng2025cweval} score models against curated insecure-code prompts
at the level of individual CWEs. Our unit of measurement differs: we score
whether a named control from a compliance framework is present in a whole
artifact, which captures omissions that leave no wrong line of code behind.

\noindent\textbf{Prompting and tuning for security.}
Tony et al.\ \cite{tony2025prompting} evaluate prompting techniques
systematically and find security-oriented prompting reduces weaknesses, and
Bruni et al.\ \cite{bruni2025forge} report a security-focused prefix cutting
vulnerabilities substantially. Fine-tuning approaches including SVEN
\cite{he2023sven} and SafeCoder \cite{he2024safecoder} pursue the same end
through training. Our single generic sentence is the lightest member of that
family, and we quantify both what it recovers and where it stops.

\noindent\textbf{Measurement reliability.}
The static-analysis literature has documented for years that developers distrust
pattern-based tools largely because of false positives \cite{johnson2013why},
that scaling detection over real code is dominated by false-positive management
\cite{bessey2010billion}, and that industrial deployment turns on controlling
that error \cite{sadowski2018google}. Tessa et al.\
\cite{tessa2026adversarial} report the complementary failure for generated code
specifically, finding that static analyzers overestimate the security of
defended generations under adversarial prompting. Our measurement section
(\S\ref{sec:measurement}) is a small addition to this line: we report how our own
text-matching scorer behaved against model-generated code, and how far the
correction ran.

\noindent\textbf{Judges as raters.}
Using a strong model to grade output is now standard practice
\cite{zheng2023judging,liu2023geval}, with documented biases
\cite{wang2024fair}. We use a blinded model grader only as an independent second
rater to test our own hand-verification for bias, and we report agreement
without treating the grader as ground truth (\S\ref{sec:method-raters}).

%------------------------------------------------------------------------------
\section{Method}
\label{sec:method}

\subsection{Design}
Three models, two prompt conditions, and four use cases: the core design is the
two models available when collection started (Fable~5, Opus~4.8) crossed with two
conditions and four use cases, giving 16 generations. Opus~5 shipped 20 days
later and was added as a third arm of 8 further generations under the conditions
recorded in \S\ref{sec:method-deviations}, for 24 in total.

The two conditions differ by one sentence:
\begin{itemize}
  \item \textbf{Round~1, neutral.} The task stated with no security or
    compliance language of any kind. This measures what the model does by
    default.
  \item \textbf{Round~2, directed.} The identical task, plus one sentence
    declaring the data sensitive and the artifact subject to SOC~2.
\end{itemize}
We call that added sentence the \textbf{SOC~2 directive} throughout. It is
appended to the user prompt, so the model holds it in context on first receipt,
and it is not placed in a system prompt, a repository file, or a tool constraint. The
directive is deliberately generic: it names the standard and the sensitivity of
the data and no individual control, which separates ``the model knows
this control and did not apply it'' from ``the model had to be told the control
by name'' (\S\ref{sec:res-blindspots}). Every other byte of prompt text is
identical between rounds, each cell was generated in a fresh session with no
prior context, and both rounds' prompts for all four use cases are published
verbatim (\S\ref{sec:availability}).

\subsection{Model selection}
\label{sec:method-models}
We tested flagship models from a single vendor, and both halves of that
decision constrain what the results support.

Holding capability at the frontier isolates the variable we care about, since a
smaller model conforming less would speak to general capability while leaving
open whether compliance behavior gets elicited, so a capability gradient would
have confounded the prompt effect we set out to measure. Restricting to one
model family controls for training data, post-training recipe, system-prompt
defaults, and product surface, any of which could otherwise explain a
cross-vendor gap. The cost is breadth: our model-comparison result is a claim
about models inside this family, and \S\ref{sec:res-models} shows it needs
narrowing even there. Cross-vendor replication is the extension we would run
first, and we treat its absence as the principal limit on generalization
(\S\ref{sec:threats}).

Model identity should be taken from the responding model, since a request can be
served by a fallback or a routed variant and the model that answers need not be
the one requested. For the first two arms we record a platform-reported
identifier with each generation. For the Opus~5 arm the identifier is only
operator- and platform-asserted, with no read-back from a response field, which
we list among that arm's deviations. A replication should capture the responding model field per
generation and store it with the artifact.

\subsection{Use cases}
Admission required that compliance decompose into concrete checkable artifacts,
10 to 15 binary items per task, with no room for a subjective judgment. Within
that constraint we chose four tasks to span a hypothesized difficulty axis:

\begin{itemize}
  \item \textbf{S3 CLI (boto3).} Imperative cloud glue, where hardening is a
    set of separate API calls that nothing in the task demands.
  \item \textbf{Authentication service (Flask or FastAPI).} SOC~2's densest
    control family, CC6.x logical access, and the place where an insecure
    generation does the most damage.
  \item \textbf{RDS Terraform.} Declarative infrastructure-as-code, added as a
    replication test after the S3 result came in low, and chosen because it adds
    two insecure constructions the first two tasks cannot exhibit: open
    security-group ingress and a hardcoded master password.
  \item \textbf{File upload holding personal data.} A blended case: the handler
    is application code, where defaults were strong, while encryption,
    retention, and audit logging are opt-in lifecycle scope, where they were
    weak. The prompt asks for ``name and email'' and never says ``PII,'' so the
    task doubles as a recognition test.
\end{itemize}

\subsection{Rubrics, and what a percentage means}
\label{sec:method-rubrics}
Each use case has a binary rubric of 12 to 15 items, written and frozen before
any output existed. Every item maps to a specific Trust Services Criterion.
Items come in two kinds: \textbf{Positive} items assert a control that must be
present, and a miss is an \emph{omission}, while \textbf{Commission} items assert
an insecure act that must be absent, and a miss means the model actively did
something dangerous. We pre-declared that commissions weigh heavier than
omissions, on the reasoning that a missing nice-to-have is a review comment
while a public bucket is an incident, and we count and report them separately,
never netting them against positive items.

SOC~2 does not prescribe technical measures: like ISO~27001 it states control
objectives and leaves the satisfying implementation to the entity and its
auditor, so no canonical checklist exists against which code can be declared
conformant, and any instrument that emits a percentage has supplied one. Ours is
no exception, since each rubric item is our mapping from a criterion to a
technical artifact we judge probative of it, and a different reader could map the
same criterion differently, weight items differently, or accept a compensating
control we scored as absent. Two consequences follow: our percentages measure
conformance to a stated technical mapping of SOC~2, so a cell at 100\% would not
thereby pass an audit and a cell at 47\% is not thereby an audit finding, and
because the mapping is ours, it is part of the contribution and has to be
inspectable, which is why every item and its criterion reference is published in
full. Readers who disagree with a mapping can rescore the unchanged artifacts
against their own.

Table~\ref{tab:tsc} lists the in-scope criteria, why each is checkable in
generated code, and one representative item per criterion, the single item being
a presentation choice for space. Several criteria carry multiple items in the same
rubric, CC6.1 being represented by default encryption, customer-managed key
choice, key rotation, credential sourcing, and password-hash strength across the
four rubrics. Even so, the items for any one criterion are a purposive sample of
what static inspection can reach and do not exhaustively decompose the criterion,
since criteria like CC6.1 have organizational and runtime dimensions no static
item touches. A figure should be read as ``satisfied this many of the statically
checkable propositions we derived for the in-scope criteria in this task.''

Criteria with no artifact in a single generated file are out of scope by
construction: CC1--CC5 (control environment, communication, risk assessment,
monitoring activities, control activities), CC6.4 (physical access),
CC7.3--CC7.5 (incident response), CC9 (vendor management), A1.3 (recovery
testing), and the Privacy P-series. The upload task involves personal data but
is scored under C1.1 and C1.2, confidentiality: the P-series concerns notice,
choice, consent, and data-subject rights, which are organizational commitments a
generated handler cannot discharge, while protecting the data and disposing of it
when no longer needed are obligations it can.

\subsection{Scoring, and the three raters}
\label{sec:method-raters}
Scoring is static: nothing was executed, no \texttt{terraform apply} was run,
and no requests were sent against any generated service, so a control artifact
is evidence of configuration, leaving runtime efficacy unmeasured.

First-pass scores came from a regular-expression heuristic, one pattern set per
rubric item, applied to the source text of each generation. It has no parser, no
type information, and no notion of a resolved configuration value, so it cannot
follow a variable to its default, separate a security-group ingress rule from an
egress rule, or distinguish a pattern being invoked from the same pattern being
rejected by a validation guard. We built it at that level deliberately, because
it is representative of how rule-based compliance scanning gets applied to
source text, and because we wanted to see what that class of tool does to
model-generated code. It is a first pass, and every failed item and every
commission flag was then hand-verified against the code, with the hand-verified
numbers canonical.

To test whether our own hand-verification was biased in the models' favor, we
added a third rater and report agreement between all three:

\begin{description}
  \item[\rone{}, pattern scorer.] The regex heuristic above, scoring all 216
    item judgments of the core design.
  \item[\rtwo{}, blind semantic grader.] An independent model-as-judge rater.
    Each of the 16 core outputs was placed in a randomized \texttt{anon-NN}
    directory stripped of model identity, round label, and all prior scores, and
    graded by a separate agent instance given only the code and the rubric.
    Graders were instructed to judge semantically: a compliant value supplied
    through a variable with a compliant default counts as present, code that
    defends against a pattern is not a commission, and a secure-by-construction
    design satisfies the corresponding item. \rtwo{} also scored all 216.
  \item[\rthree{}, human canonical.] The hand-verification, directed at every
    failed item and every commission flag.
\end{description}

\rthree{} did not cover all 216 judgments, and the asymmetry runs in the models'
favor. Because human effort was directed at the scorer's negatives, items the
scorer marked as passing were not independently re-adjudicated unless \rtwo{}
disagreed, so \rthree{} is a complete audit of the scorer's failures and a
partial audit of its passes. Some over-credits may therefore persist in the
canonical column, and the risk is demonstrable: defect X4
in \S\ref{sec:res-commissions} is a real insecure construction the scorer passed,
and it surfaced only because the Opus~5 arm re-verified passing items too. We
did not repeat that exhaustive pass over the core design, so we cannot bound how
many further over-credits it holds, and a full human pass over every judgment is
pre-committed alongside the repetition run (\S\ref{sec:threats}).

\subsection{What changed during the study, and when}
\label{sec:method-deviations}
Two facts about this study are in tension unless the sequence is explicit, so we
log it.

\emph{2026-07-05, before any generation.} The four positive-item rubrics and an
initial commission list were written and frozen. That list held four insecure
acts: a public ACL or broad bucket policy, a plaintext or weak password hash, a
hardcoded credential, and personal data written to logs. Predictions for the RDS
and upload use cases were registered before those cells existed.

\emph{2026-07-05, after scoring.} The initial list produced zero commission
failures across 16 generations, and all ten flags the scorer raised were false
positives on inspection.

\emph{2026-07-20, after that null result.} We judged the null to be a property
of the list, since the code plausibly still held insecure acts the list did not
name, and added five classes: debug-mode
disclosure, broken object-level authorization, injection (SQL and command),
unsafe deserialization, and download-path traversal. Re-scoring the unchanged
artifacts produced defects X1 through X3. This expansion was post hoc: the
positive-item rubrics were untouched and no artifact was regenerated, so the
expansion cannot have been fitted to any model's advantage, but it was informed
by having read the corpus, and X1 through X3 should be read as the product of a
list chosen after seeing the code.

\emph{2026-07-25.} The Opus~5 arm was generated and scored against the expanded
list, frozen before that arm existed, with four predictions registered before
generation. X4 comes from this arm and is the one commission finding produced
under a genuinely pre-declared list.

The list is an instrument applied to finished artifacts and appears in no
prompt, so expanding it changed the measurement while leaving the outputs untouched, and
the defects were in the files the whole time.

The added items did not come from the standard: the Trust Services Criteria name
control objectives, so CC6.3 requires access to be authorized by role and least
privilege and does not enumerate ``an unauthenticated download endpoint.'' Every
commission item is therefore an interpretation, and these five were taken from
the prohibition list of the control-to-rule guardrail described in
\S\ref{sec:conclusion}, which is generated from an organization's live SOC~2
control set and maps each rule back to the control it enforces. That interpretation
is organizational and carries no normative force.

%------------------------------------------------------------------------------
\section{Assumptions}
\label{sec:assumptions}

Five assumptions carry the results, and we state each with how far it is defended.

\textbf{A1. Rubric items are valid proxies for their criteria.} Defended by
publishing every mapping and its criterion reference, and limited by
\S\ref{sec:method-rubrics}: a satisfied proxy is evidence for conformance with
the underlying control without proving it.

\textbf{A2. Static inspection captures the compliance-relevant behavior of these
artifacts.} Defended by task selection, since all four use cases were admitted
on the basis that their controls are visible in source. Runtime behavior such as
actual TLS negotiation is out of scope.

\textbf{A3. The commission list is complete enough that a null result would
mean something.} This assumption failed, which is why
\S\ref{sec:method-deviations} exists and why we now treat commission counts as
lower bounds. Even the expanded list is not provably exhaustive.

\textbf{A4. One generation per cell represents the model's behavior on the
task.} Not defensible for fine differences, and no claim here rests on a
one-item or two-item gap. A repetition run is the first follow-on
(\S\ref{sec:threats}).

\textbf{A5. The neutral prompt is genuinely neutral.} Defended by publishing
both prompts verbatim so a reader can check that Round~1 contains no security
language. A separate question is whether a model reads an empty working directory
as a demonstration to complete, with production stakes absent, which we treat as
a threat in \S\ref{sec:threats}.

\begin{table*}[t]
\centering
\small
\caption{In-scope SOC~2 Trust Services Criteria, why each is checkable in
generated code, and one representative rubric item per criterion. Criteria with
no single-file artifact (CC1--CC5, CC6.4 physical access, CC7.3--CC7.5 incident
response, CC9 vendor management, A1.3 recovery testing, and the Privacy
P-series) are out of scope by construction (\S\ref{sec:method-rubrics}).}
\label{tab:tsc}
\begin{tabular}{@{}p{1.8cm} p{5.0cm} p{6.4cm} p{2.4cm}@{}}
\toprule
\textbf{Criterion} & \textbf{What it requires} & \textbf{Representative rubric item} & \textbf{Use case(s)} \\
\midrule
CC6.1 & Logical access: encryption, authentication, credential and key protection & Default bucket encryption; bcrypt/argon2 hashing; \texttt{storage\_encrypted}; no hardcoded secret & all four \\
CC6.2 & Register and authorize users; disable credentials when no longer valid & Account deprovisioning path; single-use expiring reset token & auth \\
CC6.3 & Role-based access, least privilege, segregation of duties & No public ACL; download restricted to the owning user & S3, upload \\
CC6.6 & Protection from threats outside the system boundary & S3 Public Access Block; no \texttt{0.0.0.0/0} ingress; upload size limit & S3, RDS, upload \\
CC6.7 & Restrict and encrypt data in transmission & TLS-only bucket policy; \texttt{rds.force\_ssl}; Secure/HttpOnly cookies & S3, RDS, auth \\
CC6.8 & Prevent or detect malicious software & Malware-scan note on uploads; no unsafe deserialization; dependency pinning & upload, all four \\
CC7.1 & Detect configuration changes and vulnerabilities & Provider version pinning; no known-vulnerable dependency & RDS, all four \\
CC7.2 & Monitor components for anomalies and security events & S3 access logging; CloudWatch log export; auth-event audit logging & all four \\
CC8.1 & Change management and configuration integrity & \texttt{deletion\_protection}; no \texttt{skip\_final\_snapshot}; safe collision handling & S3, RDS \\
C1.1/C1.2 & Identify, protect, and dispose of confidential information & No personal data in logs; encryption at rest for user files; retention and deletion & auth, upload \\
A1.1/A1.2 & Capacity, backups, and recovery & Upload size limit; \texttt{backup\_retention\_period}; versioning; \texttt{multi\_az} & S3, RDS, upload \\
PI1.2 & Completeness and accuracy of system inputs & File-type allowlist; server-side content verification; input validation & auth, upload \\
\bottomrule
\end{tabular}
\end{table*}

%------------------------------------------------------------------------------
\section{Results}
\label{sec:results}

\subsection{Unprompted conformance, and what predicts it}
\label{sec:res-unprompted}
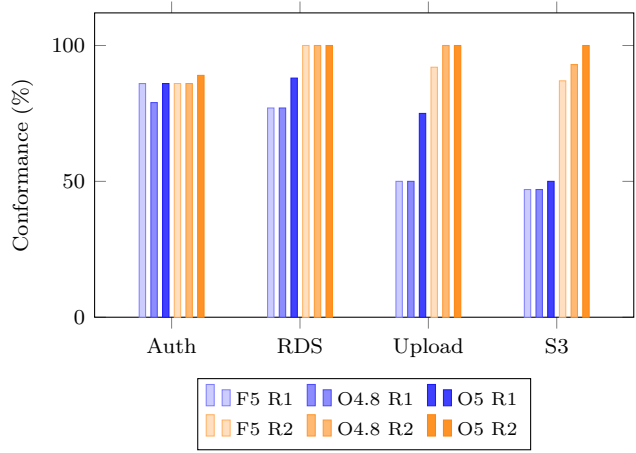
\begin{figure}[t]
\centering
\begin{tikzpicture}
\begin{axis}[
  width=\columnwidth, height=5.6cm,
  ybar, bar width=2.4pt,
  ymin=0, ymax=112,
  ylabel={Conformance (\%)},
  symbolic x coords={Auth, RDS, Upload, S3},
  xtick=data,
  enlarge x limits=0.20,
  legend style={at={(0.5,-0.20)}, anchor=north, legend columns=3,
    font=\scriptsize, /tikz/every even column/.append style={column sep=3pt}},
  tick label style={font=\footnotesize},
  label style={font=\footnotesize},
]
\addplot+[fill=blue!20, draw=blue!50] coordinates {(Auth,86)(RDS,77)(Upload,50)(S3,47)};
\addplot+[fill=blue!45, draw=blue!70] coordinates {(Auth,79)(RDS,77)(Upload,50)(S3,47)};
\addplot+[fill=blue!75, draw=blue!90] coordinates {(Auth,86)(RDS,88)(Upload,75)(S3,50)};
\addplot+[fill=orange!25, draw=orange!60] coordinates {(Auth,86)(RDS,100)(Upload,92)(S3,87)};
\addplot+[fill=orange!55, draw=orange!80] coordinates {(Auth,86)(RDS,100)(Upload,100)(S3,93)};
\addplot+[fill=orange!85, draw=orange!95] coordinates {(Auth,89)(RDS,100)(Upload,100)(S3,100)};
\legend{F5 R1, O4.8 R1, O5 R1, F5 R2, O4.8 R2, O5 R2}
\end{axis}
\end{tikzpicture}
\caption{Conformance by use case and condition across three model arms. Neutral
prompts in blue, directed in orange, with the darkest bar of each shade being
Opus~5. Within a shade the bars are close together, which is the model result;
between shades they are far apart, which is the prompt result.}
\label{fig:bars}
\end{figure}

With no security language in the prompt, conformance ran from 47\% to 88\%
depending on the task (Table~\ref{tab:canonical}, blue bars in
Fig.~\ref{fig:bars}). The spread is the first result, and the ordering inside it
refuted the hypothesis we registered.

We predicted that application-security defaults would be strong, infrastructure
provisioning defaults weak, and data-lifecycle defaults weakest, on the
reasoning that secure auth patterns saturate the training corpus while compliant
infrastructure is invisible extra scope. The authentication service behaved as
predicted at 79--86\%, and the RDS Terraform module did not: we registered a
40--60\% band for it and it came in at 77\% for both original models, 30 points
above the boto3 script doing comparable provisioning work.

The variable that sorts the data is whether a control is part of how the code is
normally written. Password hashing is how you write a login endpoint, so bcrypt
and argon2 appear without being asked for. The \texttt{aws\_db\_instance}
resource has \texttt{storage\_encrypted}, \texttt{backup\_retention\_period}, and
\texttt{deletion\_protection} as enumerable attributes in the developer's field
of view, and all three appeared unprompted. Compare \texttt{boto3}, where
\texttt{create\_bucket} plus \texttt{upload\_file} satisfies the task completely
and hardening requires four further API calls that nothing in the request
demands and whose absence raises no error. Controls sitting in that second
category, including S3 hardening calls, encryption at rest for uploaded files,
retention, MFA hooks, magic-byte verification, and malware scanning, appeared at
47--75\%.

Two corrections to the low end of that range both shrink the effect we
are describing. First, part of the S3 deficit reflects redundant code where the
protection already exists. AWS has applied SSE-S3 to all new buckets by default
since January 2023 and enabled Block Public Access and disabled ACLs on new
buckets since April 2023, so an uploader that never calls
\texttt{put\_bucket\_encryption} still writes encrypted objects and one that
never calls \texttt{put\_public\_access\_block} still lands on a bucket that
blocks public access. Crediting both items as satisfied in effect moves S3
Round~1 from 7/15 (47\%) to 9/15 (60\%) for the two original arms, which reduces
the S3-versus-RDS gap from 30 points to 17. The controls that remain missing in
effect once the defaults are credited are the TLS-only bucket policy, versioning,
access logging, and lifecycle rules. Second, the idiom account is a reading of
four data points arrived at after seeing them, so it is a hypothesis this study
generated and did not test.

\begin{table}[t]
\centering
\small
\setlength{\tabcolsep}{3pt}
\caption{Hand-verified conformance, as fraction of rubric items satisfied.
R1 neutral, R2 directed. F5 = Fable~5, O4.8 = Opus~4.8, O5 = Opus~5. Opus~5 is
scored under the original arms' stricter convention for inapplicable items
(\S\ref{sec:res-models}); under the alternative reading its auth R2 is 96\%.}
\label{tab:canonical}
\begin{tabular}{@{}lcccccc@{}}
\toprule
& \multicolumn{3}{c}{\textbf{Round 1 (neutral)}} & \multicolumn{3}{c}{\textbf{Round 2 (directed)}}\\
\cmidrule(lr){2-4}\cmidrule(lr){5-7}
\textbf{Use case} & F5 & O4.8 & O5 & F5 & O4.8 & O5\\
\midrule
Auth service  & 86\% & 79\% & 86\% & 86\%  & 86\%  & 89\%\\
RDS Terraform & 77\% & 77\% & 88\% & 100\% & 100\% & 100\%\\
Upload (PII)  & 50\% & 50\% & 75\% & 92\%  & 100\% & 100\%\\
S3 CLI        & 47\% & 47\% & 50\% & 87\%  & 93\%  & 100\%\\
\bottomrule
\end{tabular}
\end{table}

\subsection{What one sentence buys}
\label{sec:res-directive}
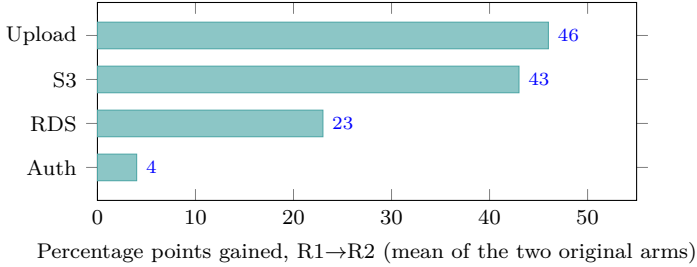
\begin{figure}[t]
\centering
\begin{tikzpicture}
\begin{axis}[
  width=\columnwidth, height=4.2cm,
  xbar, bar width=10pt,
  xmin=0, xmax=55,
  xlabel={Percentage points gained, R1$\rightarrow$R2 (mean of the two original arms)},
  symbolic y coords={Auth, RDS, S3, Upload},
  ytick=data,
  nodes near coords, nodes near coords align={horizontal},
  nodes near coords style={font=\scriptsize},
  tick label style={font=\footnotesize},
  label style={font=\footnotesize},
  enlarge y limits=0.25,
]
\addplot+[fill=teal!45, draw=teal!70] coordinates {(4,Auth) (23,RDS) (43,S3) (46,Upload)};
\end{axis}
\end{tikzpicture}
\caption{The directive buys most where unprompted defaults are weakest. Auth was
already at 82\% and had the least room, which is also where the ceiling of
\S\ref{sec:res-blindspots} appears.}
\label{fig:effect}
\end{figure}

The SOC~2 directive moved every cell to 86--100\%, worth 23 to 50 percentage
points, and it bought the most where unprompted defaults were weakest
(Table~\ref{tab:elicitation}, Fig.~\ref{fig:effect}). Averaged over the two
original arms the gains were 43 points on S3, 46 on upload, 23 on RDS, and 4 on
auth. Opus~5's corresponding gains were 50, 25, 12, and 3, the same ordering
compressed because its neutral baselines start higher.

The mechanism is visible in the diffs. Directed S3 runs added the full hardening
suite: \texttt{put\_bucket\_encryption} with a customer-managed key,
\texttt{put\_public\_access\_block} with all four settings, versioning, a
deny-based TLS-only bucket policy, access logging, and lifecycle rules. Directed
RDS runs added the three obscure attributes all neutral runs had missed, a
customer-managed KMS key, \texttt{rds.force\_ssl}, and IAM database
authentication, and Opus~5's directed run added \texttt{enable\_key\_rotation}
with an explicit rotation period. Directed upload runs added encryption at rest,
content verification by magic bytes, per-document bearer tokens, and a retention
path.

Nothing in the neutral prompts prevented any of this, the controls were within
reach the whole time, and the directive converted them from work nobody asked
for into work the request named. This is the study's central practical finding
and it is also the reason to be careful about the word ``recovers'': the
directive raised conformance in all eight core cells, but it was measured once
per cell, so it shifts a distribution whose variance we did not measure.

\begin{table}[t]
\centering
\small
\caption{Effect of the SOC~2 directive, mean of Fable~5 and Opus~4.8. Opus~5's
corresponding gains are $+50$, $+25$, $+12$, and $+3$.}
\label{tab:elicitation}
\begin{tabular}{@{}lccc@{}}
\toprule
\textbf{Use case} & \textbf{R1} & \textbf{R2} & \textbf{$\Delta$}\\
\midrule
S3 CLI      & 47\%  & 90\%  & $+43$\\
Upload      & 50\%  & 96\%  & $+46$\\
RDS         & 77\%  & 100\% & $+23$\\
Auth        & 82\%  & 86\%  & $+4$\\
\bottomrule
\end{tabular}
\end{table}

\subsection{What the directive does not reach}
\label{sec:res-blindspots}
A generic instruction has a ceiling, and the authentication service shows it.
Auth conformance plateaued at 86\% in both rounds for Fable~5, and both original
models finished the directed round at 86\%. Two items failed in every auth run
and the directive did not recover either: an MFA hook, and
\texttt{Secure}/\texttt{HttpOnly}/\texttt{SameSite} cookie flags. Both services
issued bearer tokens and set no cookies, which arguably makes the cookie
item moot, and neither model said so.

The contrast with RDS is instructive: the three obscure Terraform attributes
missed unprompted were all recovered by the directive, because they belong to a
resource schema the model already reasons about when it writes that resource. MFA
and cookie flags sit outside the model's conception of ``write me a login
endpoint,'' and no amount of emphasis on the same generic sentence brings them
in, so they have to be named.

Collecting the controls that both original models missed unprompted gives the
list a standing instruction should enumerate:

\begin{itemize}
  \item \textbf{Cloud storage:} TLS-only bucket policy, versioning, access
    logging, lifecycle and retention rules, and a customer-managed key in place
    of the platform default.
  \item \textbf{Databases:} customer-managed KMS key, forced TLS
    (\texttt{rds.force\_ssl}), IAM database authentication, and key rotation.
  \item \textbf{Authentication:} MFA hooks, cookie security flags, and an
    account deprovisioning path, which no auth generation provided in any arm.
  \item \textbf{Data handling:} encryption at rest for user files, a retention
    and deletion story, server-side content verification, and malware scanning.
  \item \textbf{Dependencies:} exact version pinning, unpinned in six of the
    eight application-code generations, and vulnerability screening. Both
    Opus~4.8 upload runs pinned \texttt{python-multipart==0.0.9}, affected by
    CVE-2024-53981.
  \item \textbf{Terraform state:} every RDS generation satisfied the item asking
    that the master password not be hardcoded, and satisfying it does not prevent
    exposure, since Terraform writes the resolved credential into state in
    plaintext however it arrives. Four of six RDS generations used
    \texttt{manage\_master\_user\_password}, under which RDS generates and
    rotates the credential into Secrets Manager and Terraform never receives it:
    Fable~5 in both rounds, Opus~4.8 and Opus~5 in the directed round. The two
    neutral-prompt generations from Opus~4.8 and Opus~5 used a
    \texttt{random\_password} resource, which does land in state. Opus~5
    documented that exposure in a source comment without adopting the
    construction that avoids it.
\end{itemize}

\subsection{Insecure constructions, and how many you find}
\label{sec:res-commissions}
Actively insecure code was rare, and the count depends on the list you search
with. Our initial four-item list (public ACL, plaintext or weak password hash,
hardcoded credential, personal data in logs) found nothing across 16
generations, and all ten flags the scorer raised against it were false positives,
several of them flagging code that was defending against the flagged pattern. We
treated that null as a property of the list, since the code plausibly held acts
the list did not name, and expanded it (\S\ref{sec:method-deviations}). Re-scoring the same unchanged files produced
Table~\ref{tab:commissions}.

\begin{table}[t]
\centering
\small
\setlength{\tabcolsep}{4pt}
\caption{Insecure constructions found only after expanding the search list. All
are in neutral-prompt code, and none survives into the corresponding directed
round. The last column gives \texttt{file:line} in the cited generation.}
\label{tab:commissions}
\begin{tabular}{@{}llp{3.0cm}l@{}}
\toprule
\textbf{ID} & \textbf{Generation} & \textbf{Construction (criterion)} & \textbf{Location}\\
\midrule
X1 & Fable~5, upload R1 & Werkzeug interactive debugger reachable: remote code execution (CC6.1/CC6.6) & \texttt{app.py:176}\\
X2 & Fable~5, upload R1 & Download endpoint with no authentication or ownership check (CC6.3) & \texttt{app.py:158}\\
X3 & Opus~4.8, upload R1 & Unauthenticated download, plus \texttt{/profiles} returning every name and email (CC6.3/C1.1) & \texttt{app.py:168,199}\\
X4 & Opus~5, RDS R1 & \texttt{skip\_final\_snapshot} resolves true outside production (A1.2/CC8.1) & \texttt{main.tf:288,13}\\
\bottomrule
\end{tabular}
\end{table}

X1 is the most severe defect in the corpus and we initially classified it too
mildly. Flask's \texttt{debug=True} mounts the Werkzeug interactive debugger,
which exposes an evaluation console on unhandled exceptions and permits arbitrary
code execution against the process, so on an internet-facing handler this reaches
remote code execution well past information disclosure, and it belongs under
CC6.1 and CC6.6, having first been classified under CC7.2 monitoring.

X4 is the most informative: \texttt{main.tf:288} sets
\texttt{skip\_final\_snapshot = local.skip\_final\_snapshot}, and
\texttt{main.tf:13} resolves that to \texttt{!local.is\_prod}; with
\texttt{environment} defaulting to \texttt{"dev"}, a default plan renders
\texttt{true}, so a teardown discards the database with no final snapshot.
Production is safe and the model documents the behavior. Our scorer marked the
item \textbf{passing}, because the non-compliance is expressed through a
conditional and never written as a literal \texttt{= true}. We return to that in
\S\ref{sec:measurement}.

Two negative results sit alongside these. We registered roughly
even odds that at least one upload run would log a user's email address, the
idiomatic and genuinely reportable \texttt{logger.info(f"upload from \{email\}")}
failure. It never happened: all six upload generations logged document
identifiers only, verified against a captured server log in one case. And path
traversal from a user-supplied filename, the most common real defect in generated
upload handlers, appeared in no generation. Every upload run either sanitized the
filename or made it irrelevant by storing under a server-generated identifier and
treating the client-supplied name as display metadata, with the Opus~5 runs
additionally rejecting separators and re-checking containment after path
resolution.

\subsection{Model choice}
\label{sec:res-models}
Across the two same-generation models, no cell differs by more than one rubric
item, with complementary single-item edges: Fable~5 wrote audit logging
unprompted in auth, Opus~4.8 added lifecycle rules in directed S3 and a malware
note in directed upload. The blinded grader saw the same picture, with a maximum
gap of two items in any cell, Opus~4.8 marginally ahead in three cells, Fable~5
in none, and five cells tied.

At one generation per cell, one rubric item is the resolution floor, so this
shows an absence of detectable difference and does not demonstrate equivalence.
We cannot distinguish ``identical'' from ``differs by one'' with this design, and
we make no equivalence claim.

Opus~5 bounds the result further: it matches the older models where their
defaults were already strong (auth 86\%, S3 50\% against 47\%) and comes in
clearly higher where they were weakest, at 75\% against 50\% on upload and 88\%
against 77\% on RDS. Of the four pre-registered predictions for that arm, the
directed-band prediction was confirmed at 89--100\%, the auth-ceiling prediction
was partly refuted since the plateau moved while MFA stayed absent, the
upload-commission prediction was refuted since Opus~5 reproduced neither X1 nor
X2, and the model-invariance prediction was neither confirmed nor strictly
refuted: Opus~5 falls within one item in only two of four use cases, while only
upload exceeds the two-item threshold the refutation condition required.

Read together with the same-generation null, the model result is that compliance
conformance carries some generational signal and less than the prompt does. The
directive is worth 23 to 50 points; the largest cross-generation difference we
measured is 25.

One scoring convention affects this comparison and we report the stricter
reading throughout. Opus~5's verifier judged the cookie-flag item inapplicable
because the service issues bearer tokens and sets no cookie, whereas the original
arms scored that same situation as a failure. Under the original convention
Opus~5's directed auth score is 12.5/14 (89\%), which is the figure in
Table~\ref{tab:canonical}; under the alternative it is 13.5/14 (96\%). Either
reading clears the 86\% plateau, and pooling the three arms would require settling
that convention first, which is one reason we do not pool them.

%------------------------------------------------------------------------------
\section{Measurement reliability}
\label{sec:measurement}

Every number in \S\ref{sec:results} is hand-verified, and this section explains
why that was necessary. A regular-expression scanner being poor at
understanding code is not a discovery, but the size and shape of its error
determined our own first-pass numbers and matter because teams gate compliance
on tools of this class.

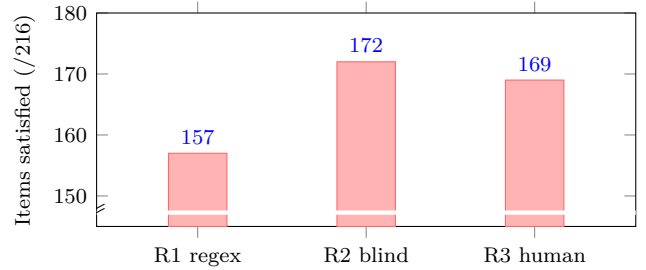
\begin{figure}[t]
\centering
\begin{tikzpicture}
\begin{axis}[
  width=\columnwidth, height=4.4cm,
  ybar, bar width=22pt,
  ymin=145, ymax=180,
  ylabel={Items satisfied (/216)},
  symbolic x coords={{R1 regex}, {R2 blind}, {R3 human}},
  xtick=data,
  nodes near coords, nodes near coords style={font=\footnotesize},
  enlarge x limits=0.30,
  tick label style={font=\footnotesize},
  label style={font=\footnotesize},
]
\addplot+[fill=red!30, draw=red!60] coordinates {({R1 regex},157) ({R2 blind},172) ({R3 human},169)};
\draw[white, line width=1.6pt]
  ([yshift=5.2pt]current axis.south west) -- ([yshift=5.2pt]current axis.south east);
\draw[black, line width=0.4pt]
  ([xshift=-3pt,yshift=3.6pt]current axis.south west) -- ([xshift=3pt,yshift=6.8pt]current axis.south west);
\draw[black, line width=0.4pt]
  ([xshift=-3pt,yshift=5.2pt]current axis.south west) -- ([xshift=3pt,yshift=8.4pt]current axis.south west);
\end{axis}
\end{tikzpicture}
\caption{Rater totals. \textbf{The axis is broken:} it starts at 145, marked at
the origin, so the bars show differences between raters and cannot be read as
proportions of 216. The blinded grader lands close to the human total and both
sit above the pattern scorer.}
\label{fig:raters}
\end{figure}

Table~\ref{tab:triangulation} gives the three-rater comparison over the 216 item
judgments of the core design, and Fig.~\ref{fig:raters} plots the totals. The
scorer \rone{} totalled 157 against the human 169, a mean signed error of
$-0.75$ per cell, and it never exceeded the human score in any cell. The blinded
semantic grader \rtwo{} totalled 172, erring in both directions, with a mean
signed error of $+0.19$. At item level \rone{} and \rtwo{} agreed on 87.5\% of
judgments (Cohen's $\kappa=0.658$ \cite{cohen1960kappa}), and 21 of their 27 disagreements were cases
where the scorer scored lower.

The corrections cluster where better engineering defeats literal matching.
Parameterization reads as non-compliance: \texttt{backup\_retention\_period =
var.backup\_retention\_days} with a default of 30 and a validation floor of 7
fails a literal-value pattern. Defensive code matches the attack it prevents: a
Terraform \texttt{validation} block whose condition rejects
\texttt{0.0.0.0/0} in an ingress list was flagged as open ingress, as was
README prose describing that rejection, while the only literal
\texttt{0.0.0.0/0} in the configuration was an egress rule. Secure-by-
construction designs make checks moot: storing uploads under a server-generated
UUID eliminates path traversal without ever calling
\texttt{secure\_filename}, so the item that looks for the sanitizer fails. One
generation shipped a vendored \texttt{.venv/}, and site-packages source produced
four flags at once. A hash-algorithm pattern fired on \texttt{sha256} inside a
\texttt{token\_digest()} helper hashing a 256-bit CSPRNG session token, where no
stretching is needed, while the passwords in the same file were scrypt.

The error does not run one way, which is the part that matters for anyone
building this kind of check. X4 in \S\ref{sec:res-commissions} is a real
insecure construction the scorer passed, because the non-compliant value was
reached through \texttt{!local.is\_prod} and never written as a literal. Our
audit was also asymmetric: human effort went to the scorer's failures, so its
passes were not exhaustively re-checked (\S\ref{sec:method-raters}). The
$-0.75$ deficit therefore bounds total error from below without estimating it,
and the over-credit rate across the core design is unknown.

Stated at the level the evidence supports: text-level scoring of generated code
cannot resolve a configuration value, so it misses compliant configuration
expressed through a variable and non-compliant configuration expressed through a
conditional. Which direction dominates depends on which of the two happens to be
the literal. Our scorer is a deliberately simple instance and we did not test
production policy engines, so this is a caveat about a class of technique, with
no measured claim about Checkov, tfsec, or OPA \cite{checkov,tfsec,opa}. It is
also the reason our protocol requires plan-time or execution-time verification
plus human adjudication for anything reported as a result.

Two practices follow for replications: resolve values before scoring, using
\texttt{terraform plan} output or an AST in place of source text, and audit the
scanner's passes as carefully as its failures, since X4 was invisible to the
protocol until we did.

\begin{table}[t]
\centering
\small
\caption{Three-rater comparison over the 216 item judgments of the core design.
\rone{} is the pattern scorer, \rtwo{} the blinded semantic grader, \rthree{} the
hand-verification. The count is pre-expansion (\S\ref{sec:method-deviations}), so
\rtwo{} never scored X1--X3, which carry no agreement statistic and are reported
as evidenced existence claims instead.}
\label{tab:triangulation}
\begin{tabular}{@{}lccc@{}}
\toprule
& \textbf{\rone{} regex} & \textbf{\rtwo{} blind} & \textbf{\rthree{} human}\\
\midrule
Total (/216)          & 157      & 172      & 169 (ref)\\
Mean signed err/cell  & $-0.75$  & $+0.19$  & n/a\\
Exceeds \rthree{}?    & never    & sometimes& n/a\\
\midrule
\multicolumn{4}{@{}l}{\emph{Item level, \rone{} against \rtwo{}:}}\\
Agreement             & \multicolumn{3}{c}{87.5\%}\\
Cohen's $\kappa$      & \multicolumn{3}{c}{0.658}\\
Disagreements         & \multicolumn{3}{c}{27 (21 regex lower, 6 higher)}\\
\bottomrule
\end{tabular}
\end{table}

%------------------------------------------------------------------------------
\section{Threats to Validity}
\label{sec:threats}

\noindent\textbf{Sample size and non-determinism.}
This is the study's largest limitation. Each of the 24 cells is a single
generation from a non-deterministic system, so we cannot estimate within-cell
variance, differences of one to two rubric items are indistinguishable from
sampling noise, and a resampled corpus would not reproduce these artifacts. We
make no inferential claim and no equivalence claim anywhere.

What the design supports is uneven across the results, and separating them is
more useful than a blanket caveat. The directive effect of 23 to 50 points is
three to seven rubric items and sits comfortably outside single-sample noise, so
its direction and rough magnitude hold while the exact percentages should be read
as one draw. The commission findings are categorical and cited to
\texttt{file:line}, so a debug console either is or is not reachable in the
artifact we scored, though their frequency under resampling is unknown, which is
why we report them as existence claims, with no rate attached. Not supported: any
fine-grained model ranking, and the $+1.5$-item Opus~5 gaps, of which only the
$+3$-item upload gap clears the noise floor.

The remedy is more samples and we pre-commit to three generations per cell with
per-cell variance, plus a paraphrase-sensitivity run. The binding cost is
adjudication more than generation: the protocol requires hand-verification of
every failure and every flag, and the blind-grade design would grow from 216 to
648 expert judgments.

\noindent\textbf{One prompt phrasing.}
The directive effect rests on a single neutral and a single directed wording, so
the magnitude could be sensitive to phrasing. Several semantically equivalent
directed phrasings are part of the pre-committed follow-on.

\noindent\textbf{Rubric and checklist validity.}
The rubrics are a defensible and non-exhaustive operationalization of SOC~2
(\S\ref{sec:method-rubrics}), and \S\ref{sec:method-deviations} shows checklist
scope is a first-order variable: a null commission result is only as strong as the
list of acts enumerated. We mitigate by mapping every item to a named criterion,
publishing the rubrics, freezing them before generation, and reporting results
under both the original and the expanded list. Scores measure control presence in
static code, leaving runtime efficacy unmeasured.

\noindent\textbf{The neutral condition may read as a demonstration.}
Every cell was generated into an empty working directory with no repository
history, no code to imitate, and no agent-instruction file such as
\texttt{CLAUDE.md} or \texttt{AGENTS.md}. That is a legitimate measurement of
unprompted defaults and it is also a recognizable kind of context, one that
plausibly reads as greenfield work, far from a change to a production service
under compliance obligations. If so, part of Round~1 measures the model's
inference about the situation, and the neutral baseline understates what the same
model would produce inside a mature repository whose surrounding code already
carries compliance signal. The effect would fall hardest on our lowest scores,
since a repository with hardened neighbours supplies by imitation much of what
the S3 and upload runs omitted. Two manipulations would separate this and we ran
neither: generate into a repository containing hardened infrastructure code, and
generate with a compliance-bearing instruction file present but no directive in
the prompt.

\noindent\textbf{Compensating controls in a real deployment.}
We score the artifact as authored, before any deployment. Regulated
environments run preventive controls that generated configuration must pass
before it takes effect, including service control policies, account
configuration rules, plan-time policy-as-code \cite{checkov,tfsec,opa},
admission controllers, and CI gates, and those do not stop applying because the
author is a model. Rubric non-conformance should be read as what the model failed
to author, leaving open what would reach production.

That protection is unevenly distributed, and it is weakest where our insecure
constructions fell. Configuration-layer defects are largely blockable before they
take effect: unencrypted storage, absent public-access blocking, open ingress, and
X4's \texttt{skip\_final\_snapshot} are all attributes a plan-time engine can
deny. Application-layer defects are not, since no service control policy prevents
a handler from mounting a debug console (X1), serving a document without an
ownership check (X2, X3), writing personal data to a log, or shipping an unpinned
dependency. Three of our four findings are application-layer, and the one a
platform guard would have caught is also the one our scorer passed.

\noindent\textbf{Single vendor, single tier.}
All three models come from one family by design (\S\ref{sec:method-models}), and
this is the largest limit on generalization. The idiom account and the elicitation
effect could reflect this family's training distribution and post-training
recipe more than a general property of frontier code models, and neither can be
separated without a cross-vendor arm. Sampling only the top tier also means we
cannot say whether the idiom account holds for smaller models, where a plausible
alternative is that they miss idiom-embedded controls too.

\noindent\textbf{Cross-arm comparability.}
The Opus~5 arm deviates from the original conditions in four ways: it was
collected 20 days later, so the same-environment same-day control does not span the
arms; its fresh-session requirement was met by isolated clean-context sessions
in place of separate interactive ones, and the orchestrating context knew the
rubrics, so contamination is reduced without being provably eliminated; its model
identity is operator-asserted, with no read-back from a response field; and it
scores one item under a different convention (\S\ref{sec:res-models}). We treat
that arm as bounding the model result and do not pool the three arms.

%------------------------------------------------------------------------------
\section{Conclusion}
\label{sec:conclusion}

When asked to build something with no mention of security, these models produced
code that runs and that an ordinary test suite would pass, and a compliance
program would still reject it. The neutral prompt omitted controls and, in three of the
sixteen original generations, produced real vulnerabilities: a reachable
debugger permitting remote code execution, and two endpoints exposing other
users' files and personal data, all scored clean by our first commission
checklist. The lesson for a compliance program is that AI-generated code needs
the same control verification as human code, because an omitted control raises
no error, passes tests, and behaves identically to a compliant one until an
auditor or an attacker looks.

The omissions followed a pattern, with conformance running 47\% to 88\% unprompted
and tracking whether a control is part of how the code is normally written: password
hashing and \texttt{storage\_encrypted} arrive unasked, while four boto3
hardening calls, a retention policy, and an MFA hook do not, because nothing in
the task demands them and their absence raises no error. One sentence naming the
standard moved every case to 86--100\% and removed every insecure construction we
found, which places the knowledge inside the model and the default behavior
outside the prompt. A stable remainder survived that sentence, including MFA
hooks, cookie flags, and account lifecycle, and those have to be named
individually.

Model choice was the least useful lever we measured. Two same-generation
frontier models finished within one rubric item in all eight comparison cells,
the resolution floor of a one-generation-per-cell design, and a
newer-generation model improved only on the two weakest cases, by less than the
directive was worth, so a team choosing a model for compliance reasons is
optimizing the wrong variable.

The measurement carries a warning for anyone auditing AI code at scale.
Expanding the list of insecure acts we searched for turned a null result into
three real defects in the sixteen unchanged files, with a fourth surfacing in
the newer arm, so the honest form of a null commission result is ``none of the
acts we enumerated,'' and every count we report is a lower bound. The
regular-expression scorer failed whenever the decisive value was not a literal:
it under-credited well-engineered code across 21 of 27 disagreements with
semantic grading, and it passed an RDS instance that discards its data on
teardown outside production, because that setting is computed by a conditional
and never written as \texttt{= true}. Audit a scanner's passes as carefully as
its failures, and resolve configuration values before scoring.

Because the gap is prompt-shaped, the remediation we ship is prompt-shaped:
\texttt{agent-guardrails/} in the artifact repository holds a \texttt{CLAUDE.md}
generated automatically from a live SOC~2 control set, pairing a
prohibition list aimed at the constructions in Table~\ref{tab:commissions} with
per-area required controls aimed at the list in \S\ref{sec:res-blindspots}, and
an audit map tracing each rule to the control it enforces. The direct test is
registered and unrun: place that file in a repository, re-run the neutral
prompts, and check whether conformance matches the directed scores with no
directive in the prompt.

Two caveats on that remedy, which we state plainly so no reader has to discover
them. A single instruction is the cheapest intervention we measured, and it is
not the most reliable one available: its influence competes with a system prompt,
post-training dispositions, surrounding code, and tool output, its salience
decays as a session grows, and context compaction can remove it altogether, which
is why the guardrail belongs in a file that gets re-read each session, beyond any
single turn of conversation. And the environment is a stronger lever than the
instruction: \texttt{boto3} making an unhardened bucket the shortest path explains
most of the gap against Terraform, so a wrapper whose bucket-creation call is
encrypted, private, and logged by default removes the failure mode without asking
a model to remember it. We did not test secure-by-default tooling, and it is the
direction this study points at most clearly.

%------------------------------------------------------------------------------
\section*{Acknowledgments}
We thank Naren Beniwal, Molly Correia, James Ford, Jordan Hamel, Anna Hix,
Swapnil Pawar, Sebastian Rojas, Zheng Tao, Edward Thomson, Eric Tierling, and
David Zhao for their contributions to this work. We also thank our colleagues on
the Vanta security and compliance teams for their input. The models evaluated in
this study were accessed through their standard interfaces; no vendor had any
role in the study design, the rubrics, the scoring, or the reporting of results.
Any errors are our own.

\section*{Availability}
\label{sec:availability}
All artifacts are in a single repository:

\begin{center}
\url{https://github.com/OpenVanta/fable-vs-opus-compliance}
\end{center}

\noindent It holds the four rubrics with per-item criterion mappings
(\texttt{RUBRIC*.md}), the coverage and scoring-method analysis
(\texttt{RUBRIC-GAPS.md}), the pre-registered predictions for the RDS and upload
use cases (\texttt{PREDICTIONS.md}) and for the Opus~5 arm
(\texttt{RUNBOOK-opus5.md}), the verbatim Round~1 and Round~2 prompts
(\texttt{PROMPT*.md}), all 24 generations exactly as emitted
(\texttt{fable-output/}, \texttt{opus-output/}, \texttt{opus5-output/}), the
pattern scorer (\texttt{score.py}, \texttt{compare.py}), the canonical
hand-verified scores with per-item corrections (\texttt{RESULTS.md},
\texttt{RESULTS-opus5.md}), and the generated guardrail
(\texttt{agent-guardrails/}). Each generation folder holds the model's output
unmodified, including READMEs and dependency manifests, so every
\texttt{file:line} citation above resolves directly.

\end{document}